%% file: ndss2025.tex
\documentclass[conference]{IEEEtran}

\ifCLASSINFOpdf
\else
\fi

\usepackage{url}
\usepackage{natbib}
\usepackage{tikz}
\usepackage{graphicx}
\graphicspath{{images/}}
\usepackage{amsmath}
\usepackage{array}
\usepackage{amsfonts}
\usepackage[linesnumbered,vlined,ruled]{algorithm2e}
\usepackage{comment}
\usepackage{wasysym}
\usepackage{caption}
\usepackage{algpseudocode}

\usepackage{enumitem}
\usepackage{multirow}
\usepackage{subfigure}
\usepackage{pifont}
\usepackage{booktabs}
\usepackage{float}

\def \name{{\textsc{MLQEnabler}}\xspace}

\def \model{{CSS-MLS}\xspace}
\def \net{{EncGAN}\xspace}

\begin{document}
%
\title{\Large \bf \name: Enabling Secure Machine Learning Queries over Encrypted Database in Cloud Computing} 

\author{Xu Zhou, Haoyang Chen, Xinyu Lei \\

Department of Computer Science, Michigan Technological University, Houghton, MI, USA \\

E-mail: \{xzhou4, haoyangc, xinyulei\}@mtu.edu

%
}

\maketitle
\pagestyle{plain}

\input{sections/0-abs}

\ifCLASSOPTIONpeerreview
\begin{center} \bfseries EDICS Category: 3-BBND \end{center}
\fi
%
\IEEEpeerreviewmaketitle

\input{sections/1-intro}

\input{sections/2-problem}

\input{sections/4-design-v2}

\input{sections/5-analysis}

\input{sections/6-evaluation-v2}

\input{sections/8-conclusion}

\section*{Acknowledgment}

This work is supported by the National Science Foundation under Grant Number CNS-2153393. 

\bibliographystyle{plain}
\bibliography{ref}

\end{document}

%% file: sections/0-abs.tex
\begin{abstract}
%
%
In cloud computing, the public cloud service providers (CSPs) can provide cloud storage as the primary service while providing additional machine learning (ML)-based services by using the clients' data in storage.
This business model extends the border of cloud computing services and brings in new business growth possibilities.
Although it is promising, the model also brings in security concerns since the public commercial cloud cannot be fully trusted.
For example, the public commercial clouds may sell clients' sensitive data to the government or other companies.
To address the security concerns, an immediate solution is to require clients to encrypt their datasets before outsourcing to the cloud.
However, if a database is formally encrypted, then the database contains only pseudorandom numbers, making it impossible to
enable ML over it.
In this project, we propose \name (\textbf{ML} \textbf{Q}ueries \textbf{Enabler}) scheme to enable secure ML queries over encrypted database in cloud storage.
\name employs an index-aid approach to achieve security and ML capability simultaneously.
Our initial experiments show that \name achieves an acceptable security level while incurring only a slight ML performance degradation.
\end{abstract}

%% file: sections/1-intro.tex
\section{Introduction}\label{sec:intro}

\noindent
\textbf{Background and Motivation.}
In many real-world business models, clients' data is accessible by their service providers.
Multiple clients can enjoy the services provided by the centralized service providers.
Meanwhile, the service providers can exploit the merged data from multiple clients to deliver some extra machine learning (ML)-based services.
For example, when clients use Amazon to search and purchase products, Amazon can collect clients' historical search/purchase data to build its ML-based recommendation system.
We call the above business model as \textbf{P}rimary \textbf{S}ervice with \textbf{M}achine \textbf{L}earning-based \textbf{S}ervice (PS-MLS) model.
In the above example, the primary service is the online shipping provided by Amazon, the ML-based service is the product recommendation.
PS-MLS model is quite common in practice as evidenced by the fact that most newly installed apps would pop up a window to ask for the permission to access some users' private data.
Once permitted, the app-collected data can be used to provide ML-based services.

As an instance of PS-MLS model, in cloud computing, the cloud storage service (CSS) can be treated as the primary service while the cloud service provider (CSP) can also offer extra ML-based services by leveraging the massive datasets stored in the cloud.
For brevity, this business model is named \model model in this paper.
Compared with local data storage, CSS has lower costs, better performance, and higher flexibility.
As a result, CSS is increasingly prevalent and the cloud-host data volumes grow exponentially.
According to ComputerWeekly \cite{data_volume}, the volume of data on earth will increase to 175 ZB (1 ZB $\!\!=\!\! 10^{12}$ GB) by 2025 and half of data on earth will be stored in public clouds.
Due to the huge volume, even a small portions of cloud-host data is sufficient to train good ML models.
Therefore, enabling ML services in CSS is a natural and pressing demand.
%
%
\model model captures the spirit of the data sharing economy in the big data era and is expected to continue to expand in the future.

Promising as it is, \model model also raises security threats towards clients' outsourced data because the public cloud cannot be fully trusted.
First, there are financial incentives for CSPs to sell clients' personal data to the government or other companies.
In addition, some corrupted cloud employees with administrator privileges may directly peek at and steal clients' data.
Moreover, cloud server may also be vulnerable to various hacking attacks (e.g., Capital One data breach \cite{lu2019assessing}).
To address the security concerns rooted in cloud storage, clients are usually required to encrypt their datasets before outsourcing to a public cloud.
In European Union (EU), such a requirement has been enforced by the General Data Protection Regulation (GDPR) since 2018 \cite{voigt2017eu}.
However, if the clients encrypt their data before sending to the cloud, then it is difficult to offer ML service over encrypted data (i.e., the data utility is harmed).
If tens of ZB cloud-host data cannot be used, it would lead to huge economic losses for the cloud storage industry.
Thus, it is highly needed to devise a solution to maintain the ML capability over the cloud-host encrypted data while still providing data privacy protection against the untrusted cloud.

\noindent
\textbf{Limitations of Prior Art.}
To address the above problem, three types of prior techniques can be used.
The three techniques suffer from some limitations as analyzed below.
The first one is full homomorphic encryption (FHE) \cite{gentry2009fully,brakerski2011fully}.
FHE allows direct computation over the encrypted data.
%
However, the current FHE schemes are still inefficient for neural networks training, especially for deep neural networks.
For example, an FHE-based solution take several hours to train a 4-layer neural network model on a normal PC and deeper neural network training cannot be handled \cite{hesamifard2019deep}.
The second technique is differential privacy (DP) \cite{dwork2006differential,dwork2008differential}.
DP-based schemes rely on incorporating random noise into the original data or ML models.
DP-based schemes achieve a coarse-level noise addition.
If a large noise is added to the data, then the trained ML suffers from a low prediction accuracy.
Whereas if a small noise is added, then the data security strength is low.
Hence, the accuracy-precision tradeoff using DP technique is not very satisfactory.
The third technique is federated learning (FL) \cite{mcmahan2017communication,yang2019federated,kairouz2019advances}.
It enables to train an ML model with the help of a central cloud server while keeping training data distributed over multiple clients.
The major problem of FL is: it cannot provide query privacy and model privacy.
For example, the trained ML model is directly leaked to the adversary.
The trained ML model represents the intellectual property (IP) of the data owners and the IP leakage may lead to IP infringement. 
For example, the adversary may use the well-trained ML model to gain improper commercial benefits. 
Thus, protecting model privacy against the untrusted cloud is also desirable.

\noindent
\textbf{Our Approach.}
At first glance, it is a non-trivial task to devise the solution because there is a dilemma between achieving ML capability and preserving data security.
On the one hand, if clients' data is formally encrypted (e.g., using AES or 3DES \cite{stallings2006cryptography})), then the ciphertext are pseudo-random numbers, making it is impossible to be used for ML purposes.
On the other hand, if clients' data is not encrypted, then data security is breached.
In this paper, we propose \name (\textbf{ML} \textbf{Q}ueries \textbf{Enabler}) scheme to enable ML queries over encrypted databases in cloud computing.
At a high level, \name employs an index-aid approach.
In \name, each data item (i.e., a training example) in a dataset is formally encrypted (e.g., using AES/DES) to achieve strong ciphertext privacy.
For each data item, \name generates a secure index item.
The generated index items can support ML over them.
By this approach, \name can achieve strong ciphertext privacy and ML capability simultaneously.

\noindent
\textbf{Technical Challenges and Proposed Solutions.}
In \name, two major technical challenges need to be further addressed.

\begin{figure}[htbp]
\centering
\includegraphics[width=0.95\columnwidth]{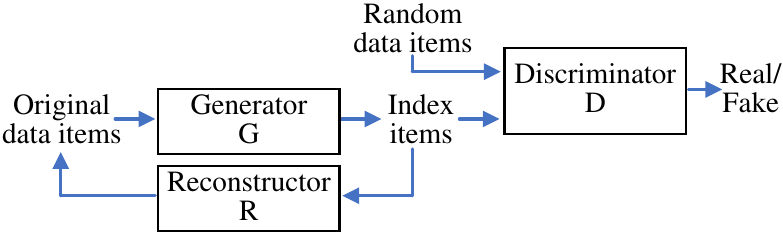}
\caption{How to generate secure index items by using \net.}
\label{fig:cyclegan}
\end{figure}

\noindent\underline{Challenge 1:} it is challenging to achieve index privacy in \name.
Since \name introduces extra index items, it should be ensured that the adversary is hard to learn useful information from the generated index items (i.e., it is hard to achieve index privacy).
To achieve index privacy, we use a generative adversarial network (GAN) to train a generator  $G$ to be used for secure index items generation.
As shown in Fig. \ref{fig:cyclegan}, on input an original data item, the generator $G$ learns to generate an index item that is indistinguishable from a randomly generated data item.
Based on cryptography theory \cite{katz2020introduction}, if the generated index item is indistinguishable from a random one, then such index generation method can achieve \emph{ciphertext-only security}.
This is because an adversary cannot learn useful information by viewing only the index items consisting of random numbers.
In \name, the well-trained generator $G$ is used for secure index items generation to protect index privacy.

\noindent\underline{Challenge 2:} it is challenging to preserve the ML capability over the secure index items.
To address this challenge, we additionally develop a reconstructor $R$ (as shown in Fig. \ref{fig:cyclegan}), which aims to reconstruct the original data item from the secure index item.
To train $R$, \name introduces a reconstruction loss, which is defined as the difference between the reconstructed data item and the original data item.
After training, it holds that $R(G(x)) \approx x$.
It is believed that ML models can be trained over $G$-generated secure index items.
%
%
%
In this paper, the secure index generation framework shown in Fig. \ref{fig:cyclegan} is called \textbf{Enc}ryption \textbf{GAN} (\net).
%
%


\begin{table}[ht]
\normalsize
\centering
\caption{Comparison between \name and other techniques (\CIRCLE: supported, \Circle: not supported, \LEFTcircle: partial).}
{\begin{tabular}{|c|c|c|c|c|}
\hline
&FHE&DP&FL&Ours \\ \hline \hline
Ciphertext privacy&\CIRCLE&\LEFTcircle&\CIRCLE&\CIRCLE \\ \hline
Index privacy&N/A&N/A&N/A&\CIRCLE \\ \hline
Query privacy&\CIRCLE&\LEFTcircle&\Circle&\CIRCLE \\ \hline
Model privacy&\Circle&\LEFTcircle&\Circle&\CIRCLE \\ \hline
\end{tabular}}
\label{table:comparsion}
\end{table}

\noindent 
\textbf{Comparison.}
Table \ref{table:comparsion} shows the comparison between \name and other techniques.
It can be found that \name can achieve four privacy-preserving goals simultaneously, thereby outperforming prior techniques in privacy protection.

\noindent 
\textbf{Our Contributions.}
This paper makes the following key contributions.



\begin{itemize}
    \item We identify and formulate the \model model, where a cloud service provider aims to support ML-based services over clients' outsourced data while preserving data privacy against an untrusted cloud.
    \item We propose \name, an index-aided framework that encrypts each data item with standard encryption while generating secure index items to support ML queries.
    \item We design a GAN-based index generation method that makes the generated index items indistinguishable from random data, thereby protecting index privacy under ciphertext-only attacks.
    \item We introduce a reconstructor with a reconstruction loss to preserve the ML utility of the secure index items, enabling ML models to be trained over protected outsourced data.
\end{itemize}

%% file: sections/2-problem.tex
\begin{figure*}[t]
\centering
\includegraphics[width=2\columnwidth]{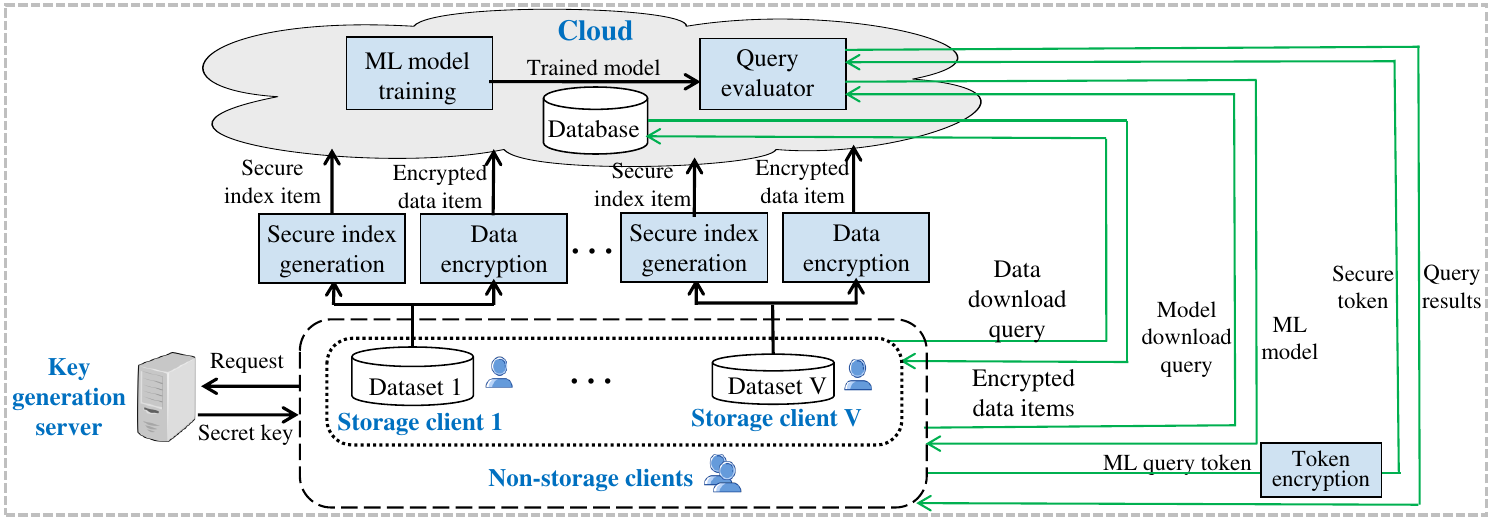}
\caption{\name system model.}
\label{system model}
\end{figure*}

\section{Problem Formulation}\label{sec:prob}
\subsection{System Model}

As depicted in Figure \ref{system model}, the proposed \name system model involves four different entities, i.e., multiple storage clients, multiple non-storage clients, a cloud server, and a key generation server (KGS).
Each storage client owns a large dataset (consisting of many data items) to be stored in the cloud server.
The storage clients expect the cloud to additionally provide ML queries processing services.
The non-storage clients do not use the cloud storage service, but they intend to use the ML queries processing services.
The cloud server is responsible for rendering data storage services as well as responding to ML queries from clients.
The KGS is a server with limited storage resources and is capable of generating secret keys for each client upon request.

In \name scheme, each storage client first formally encrypts their data items.
Then, all storage clients resort to the KGS to generate a secret key.
Next, each storage client uses the secret key to generate secure index items.
Subsequently, each storage client outsources both the encrypted data items and the secure index items to the cloud. The secure index items can be used by the cloud for ML models training.
Later, any client can send different types of queries to the cloud.
Upon receipt of the queries, the cloud can process the queries and returns the corresponding results to the client.
Since an increasing volume of data may be stored and accessible by the cloud, the cloud can periodically re-train the ML models to improve their quality (i.e., test accuracy).

\subsection{Data, ML model, and Query Types}

In this paper, we consider the most frequently used image data and neural network-based ML models.
Besides, three types queries are supported in \name. 

\begin{itemize}[leftmargin=0.35cm]
	\item \textbf{Data Download Queries.} Each storage client can send this query to selectively download data items stored in the cloud. This query is the most fundamental service provided by the cloud.
	\item \textbf{Model Download Queries.} All authorized clients can send this query to the cloud to request the trained ML model.
	\item \textbf{ML Queries.} Consider a client wants to invoke the ML model to predict a data item's label, the client can first invokes token encryption algorithm to generate a secure ML query. Then, the secure ML query can be sent to the cloud. Next, the cloud can feed the query to the trained ML model and output the ML results. Last, the ML results are returned to the client.
\end{itemize}

\subsection{Threat Model and Assumptions}
We consider the cloud server as the adversary, which is assumed to be semi-honest (i.e., honest-but-curious).
To be more specific, the cloud server is restricted to render data storage service and ML query service as specified in the protocol.
However, it also exhibits curiosity about the received data.
It may record all information it can access to learn other information that should be kept secret.
For example, the cloud server may try to infer clients' original data items from the encrypted index items.
Besides, the KGS is treated as a trusted entity.
We assume that storage client's data items are independent and identically distributed (IID).

\subsection{Design Goals}
\name should satisfy the following design goals.
\begin{itemize}[leftmargin=3mm, itemsep=2pt,topsep=0pt,parsep=0pt]

	\item \textbf{Privacy.} There are four sub-goals.

	\begin{enumerate}[leftmargin=4mm, itemsep=2pt,topsep=0pt,parsep=0pt]
	\item[1)] \textsf{Ciphertext privacy}: from the encrypted data items, it is hard for the adversary to recover the original data items.
	\item[2)] \textsf{Index privacy}: from the secure index items, it is hard for the adversary to recover the original data items.
	%

    
	\item[3)] \textsf{Token privacy}: from a secure ML query token, it is hard for the adversary to recover the original data item used to generate the query token.
    \item[4)] \textsf{Model privacy}: from the trained ML model, it is hard for the adversary to reveal the ground truth (i.e., the real ML model trained using all storage clients’ original data items).

    \end{enumerate}
    
 
	\item \textbf{Usability.} \name should ensure the query results are as accurate as possible.

\end{itemize}

%% file: sections/4-design-v2.tex




\section{\name Design}\label{sec:design}

\name consists of four major procedures: (1) key generation procedure, (2) secure index generation and data encryption procedure, (3) ML training procedure, and (4) queries processing procedure.
In this section, we elaborate on each procedure of \name.

\subsection{Key Generation Procedure}

The secret key of \name is the generator $G$ of a well-trained \net.
To train \net, KGS first collects a dataset from the Interent. 
The collected dataset is used by KGS to train $G$, $D$, and $R$ (as shown in Fig. \ref{fig:cyclegan}).
After training, the generator $G$ serves as the secret key. 
%





\noindent\textbf{Adversarial Loss.}
We adopt the adversarial loss in \cite{goodfellow2014generative} to ensure that the index items generated by $G$ are indistinguishable from the randomly generated data items, so that the privacy of index items can be protected. 
Formally, given a set of real data items $\mathcal{I} \subseteq \mathcal{X}$ ($\mathcal{X}$ denotes the data space), $G$ aims to transform each real data item $x \in \mathcal{I}$ into an index item $G(x)$ that is indistinguishable from a data item $z$ randomly sampled from $\mathcal{X}$, while $D$ aims to distinguish between $G(x)$ and $z$. 
%
%
Therefore, the adversarial loss is given by 
%
\begin{align}
    L_{\text{adv}}(G, D)
    &=
    \mathbb{E}_{z \sim \mathrm{Unif}(\mathcal{X})}[\log D(z)] \\
    &+
    \mathbb{E}_{x \sim \mathrm{Unif}(\mathcal{I})}[\log(1-D(G(x)))].
\end{align}
%
%
%


\noindent\textbf{Reconstruction Loss.}
To support ML over $G(x)$, we train $G$ jointly with a reconstructor $R$, which aims to recover $x$ from $G(x)$. 
We introduce a reconstruction loss to train $G$ and $R$. 
The reconstruction loss is defined as the distance (using the $\ell_{1}$-norm) between the reconstructed data item $R(G(x))$ and the real one $x$. 
It is given by
\begin{equation}\label{loss:cycle}
	L_{\text{rec}}(G, R)
    =
    \mathbb{E}_{x \sim \mathrm{Unif}(\mathcal{I})}[\|R(G(x))-x\|_{1}].
\end{equation}

\noindent\textbf{Full Objective.} 
To sum up, the final loss can be represented as 
\begin{equation}\label{eq:loss}
    L(G, D, R)
    =
    L_{\text{adv}}(G, D)
    +
    \lambda_{\text{rec}}L_{\text{rec}}(G, R).
\end{equation}
%
%
%
%
%
%
%
%
During the training process, \name aims to solve
\begin{equation}\label{obj:final_min_max}
G^{*}=\arg \min _{G, R} \max _{D} \mathcal{L}(G, R, D).
\end{equation}
The \net training algorithm is shown in Algorithm \ref{alg:net}.
\begin{algorithm}[htb]
	\caption{\net Training Algorithm}
	\label{alg:net}
	\LinesNumbered
    \KwIn{Generator $G$; discriminator $D$; reconstructor $R$; set of real data items $\mathcal{I}$; data space $\mathcal{X}$; total number of iterations $T$; batch size $B$.}
	\KwOut{Well-trained $G^{*}$.}
	Randomly initialize $G$, $D$, and $R$\;
	\For{iteration $t = 1:T$}{
		Draw a minibatch of $B$ real data items from $\mathcal{I}$;
        
		Draw a minibatch of $B$ random data items from $\mathcal{X}$;
		
		Compute $L(G, D, R)$ according to Eq. (\ref{eq:loss});

		Update $D$ by ascending its stochastic gradient;
		

		Update $G$ and $R$ by descending its stochastic gradient;
	}
	\Return $G^{*} = G$;
\end{algorithm}

\subsection{Secure Index Generation and Data Encryption Procedure}

After KGS finishes \net training, $G$ serves as the global secret key.
Then, KGS distributes the global secret key to each storage client.
To achieve index privacy, each storage client feed the original data items to $G$ and output the corresponding encrypted index items.
To achieve ciphertext privacy, each storage client is required to generate a local secret key for data encryption.
Then, the storage clients use their local secret keys to formally encrypt their local data items using a standard encryption algorithm (e.g, AES or 3DES).
Next, storage clients outsource both the secure index items and encrypted data items (attached with its class label information) to the cloud. 
Note that the class label information is allowed to be disclosed to the cloud.

%

\subsection{ML Training Procedure}
Upon receipt of the secure index items and their attached label information from multiple storage clients. The cloud can train different ML models directly over secure index items. 
The trained ML model is denoted as $M'$.


\subsection{ML Queries Processing Procedure}




Given a data item without an ML result (e.g., without a label), clients can query the cloud for the ML result. 
First, any clients (including both storage clients and non-storage clients) can request the global secret key $G$ from KGS. 
Then, the KGS sends the secret key $G$ to the authorized clients.
To have an ML query, the client uses $G$ to transform the data item $q$ to generate an encrypted token $G(q)$. 
Next, the encrypted token $G(q)$ is sent to the cloud. 
Upon receipt, the cloud feeds $G(q)$ to the trained ML model $M'$ to obtain the query result $M'(G(q))$. 
At last, the query result $M'(G(q))$ is returned to the client.


%

%% file: sections/5-analysis.tex
\section{\name Analysis}\label{sec:theory}

In this section, we analyze \name in terms of security and performance.

\subsection{Security Analysis}

\noindent\textbf{Ciphertext Privacy.}
In \name, each data item in clients' private datasets is formally encrypted by standard encryption methods (e.g., AES or 3DES) using clients' local secret keys. 
After being outsourced to the cloud server, an adversary is unable to restore the original data from the encrypted data item without the local secret keys.
Accordingly, ciphtertext privacy can be achieved by \name.

\noindent\textbf{Index Privacy.}
In \name, secure index items are encrypted by using $G$.
One loss function (i.e., the adversarial loss) in training $G$ is to
control output the index items to be hard to be distinguished from pseudorandom ones. 
Thus, the index privacy is protected to some extent. 
Note that there is a tradeoff between the index privacy protection strength and the ML result accuracy.

\noindent\textbf{Token Privacy.}
In \name, a client's token is encrypted by using $G$,
Hence, the token privacy is protected in the same way as the index privacy analyzed above.

\begin{figure}[htbp]
\centering
\includegraphics[width=0.85\columnwidth]{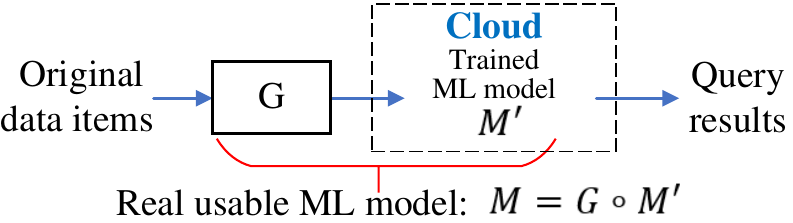}
\caption{Why model privacy is protected.}
\label{fig:model_privacy}
\end{figure}

\noindent\textbf{Model Privacy.}
In \name, the cloud is responsible for training the ML model $M'$ over secure index items.
To properly use the ML model trained on the cloud, as shown in Figure \ref{fig:model_privacy}, a client needs to first send an original data item to $G$ and then send the output to $M'$.
Therefore, the real usable ML model is $M=G\circ M'$ (i.e., $G$ has the series connection with $M'$).
Since the cloud only has access to $M'$ and the secret key $G$ is kept secret from the cloud, the cloud unable to correctly use the ML model. 
Thus, the model privacy is protected.

\noindent
\textbf{Loss-Functions-Controllable Noise Addition.}
In \name, the encrypted process by using $G$ can be viewed as a loss functions-controllable (LFC) noise-addition technique. 
Compared with differential privacy (DP), the proposed LFC noise-addition method is more fine-grained since the noise addition process in \name is directly guided by the designed loss functions via ML, whereas DP's noise addition is only determined by a pre-specific coarse noise scale parameter (defined in Laplace distribution). 
The proposed LFC noise addition method can be viewed as a new type privacy-preserving technique. 
It be used in many other applications beyond this paper.

\subsection{Performance Analysis}


\noindent
\textbf{Dataset Add/Deletion Handling.}
Each storage client may dynamically add and delete the data items in its cloud storage. %
The cloud server can selectively choose the updated dataset to retrain updated ML models.
For example, if clients add more data items in cloud storage, the cloud can re-train better ML models by harnessing more training data.

\noindent
\textbf{Security-Space Tradeoff.}
In \name, there exists a tradeoff between security and space cost.
For each data item to be stored, the client needs to generate a secure index item that is used for ML purposes, so the space cost is doubled in \name. 
That is, \name sacrifices space for ML purposes.
The extra space cost in \name is not a big issue since storage space is a very cheap resource in cloud computing.
%

%% file: sections/6-evaluation-v2.tex
\section{Experiments}
\label{sec:eval}


\subsection{Experimental Setup}

\name is applicable to various types of data. In this paper, we only apply it to image data and leave other data types (e.g., text) for future work. We conduct extensive experiments on the classification task.

\noindent\textbf{Datasets.}
In our experiments, six image datasets are adopted, including DIV2K \cite{agustsson2017ntire}, CIFAR-10 \cite{krizhevsky2009learning}, CIFAR-100 \cite{krizhevsky2009learning}, Tiny-ImageNet \cite{in200}, GTSRB \cite{stallkamp2012man}, and CelebA \cite{liu2015deep}.
We train \net on DIV2K and evaluate it on CIFAR-10, CIFAR-100, Tiny-ImageNet, GTSRB, and CelebA. 
%
The datasets are briefly described as follows.
\begin{itemize}[leftmargin=3mm, itemsep=2pt,topsep=0pt,parsep=0pt]

    \item \textbf{DIV2K}: The DIV2K dataset was introduced for example-based single image super-resolution (SR) and also used for image hiding in \cite{jing2021hinet, guan2022deepmih}. It contains 800 training images and 100 validation images with 2K resolution.
    
    \item \textbf{CIFAR-10} \cite{krizhevsky2009learning}: CIFAR-10 is an image classification dataset with ten classes and consists of 50,000 training images and 10,000 testing images with the same image size of $32 \times 32$.
    

    \item \textbf{CIFAR-100} \cite{krizhevsky2009learning}: CIFAR-100 is an image classification dataset with 100 classes and consists of 50,000 training images and 10,000 testing images with the same image size of $32 \times 32$.

    \item \textbf{Tiny-ImageNet} \cite{in200}: The Tiny-ImageNet dataset is a widely used image classification benchmark with 200 distinct classes. It contains a total of 100,000 training images and 10,000 testing images, each with a uniform resolution of $64 \times 64$ pixels.
    
    \item \textbf{GTSRB} \cite{stallkamp2012man}: The GTSRB dataset comprises 51,800 traffic sign images in 43 categories. The dataset is divided into 39,200 training images and 12,600 testing images. The image size is $32 \times 32$.
    
    \item \textbf{CelebA} \cite{liu2015deep}: The CelebA dataset is a large-scale face attributes dataset with 202,599 colored celebrity images (162,770 for training, 19,867 for validation, and 19,962 for testing).
    Each image has a size of $218 \times 178$ and 40 binary attributes.
    Following the configuration in \cite{duan2024conditional, nguyen2021wanet, salem2022dynamic}, we select the top three most balanced attributes (\textit{i.e.}, Heavy Makeup, Mouth Slightly Open, and Smiling) and concatenate them into eight categories.
    Our experiments only use its training set and testing set for consistency with other datasets.
\end{itemize}

\noindent\textbf{Model Architectures.}
For the target ML model, we adopt the CNN-based ResNet-18 \cite{he2016deep} and the transformer-based SwinV2-T \cite{liu2022swin}.

\noindent\textbf{Parameter Settings.} 
%
%
We train \net for 200 epochs on the DIV2K dataset using the Adam \cite{kingma2014adam} optimizer with $\beta_{1}=0.5$, $\beta_{2}=0.999$.
The batch size is 1 for each iteration.
The initial learning rate is set to $2 \times 10^{-4}$. 
It remains constant for the first 100 epochs and then linearly decays to $0$ over the subsequent 100 epochs. 
Data augmentation techniques include random horizontal flipping, random vertical flipping, and random cropping to $224 \times 224$ pixels without padding.

For image classification tasks, we train ResNet-18 for 200 epochs with a batch size of 128 using the SGD optimizer with a momentum of 0.9, a weight decay of 5e-3, and an initial learning rate of 0.01, which is divided by 10 after 100 and 150 epochs.
We employ the AdamW \cite{kingma2014adam} optimizer with $\beta_{1}=0.9$, $\beta_{2}=0.95$, and a weight decay of 0.1 to train SwinV2-T for 300 epochs with a batch size of 128.
The learning rate increases in the first 20 epochs from 0 to 1e-3 with a linear warm-up scheduler and then decreases in the remaining 280 epochs to 0 with a cosine decay scheduler.
Without loss of generality, all images from the classification datasets are resized to $224 \times 224$ pixels.
We use random horizontal flipping and random cropping to $224 \times 224$ pixels after a 28-pixel padding on each side of the resized images for data augmentation on the CIFAR-10, CIFAR-100, Tiny-ImageNet, and CelebA datasets.

\noindent\textbf{Implementation.} We conduct the experiments on a single NVIDIA GeForce RTX 3090 GPU using the PyTorch \cite{paszke2019pytorch} framework.

\subsection{Evaluation Metrics}
%
%

\noindent
\textbf{Patch Information Entropy (Patch IE).}
%
%
Patch IE measures the average patch-wise information entropy of an image.
Given an image of size $C \times H \times W$, it can be partitioned into a sequence of $N$ non-overlapping patches $\{I_p^i\}_{i=1}^N$, each of spatial size $P \times P$, where $N = HW/P^2$. 
Patch IE is defined as
\begin{equation}
\mathrm{PIE}_{p}(x)
=
\frac{1}{N}\sum_{i=1}^{N}H(I_p^i),
\end{equation}
where $H(I_p^i)$ computes the information entropy of the $i$-th image patch.



\noindent
\textbf{Authorized Accuracy (AA).}
AA quantifies how much an obfuscation method preserves data utility for DNN training. It is defined as the accuracy of a model trained on obfuscated data. A higher AA indicates higher training utility. Note that DP and InstaHide evaluate trained models on original testing data, while VIM and our method evaluate on obfuscated testing data.

\noindent
\textbf{Unauthorized Accuracy (UA).}
UA evaluates the effectiveness of model intellectual property protection. It is defined as the accuracy of a model trained on obfuscated data when tested directly on the plain (non-obfuscated) testing set. A high UA suggests that the trained model is usable on plain images without authorization, whereas a low UA indicates strong protection against unauthorized deployment.

\subsection{Main Results}

\begin{figure}
    \centering
    \includegraphics[width=1\linewidth]{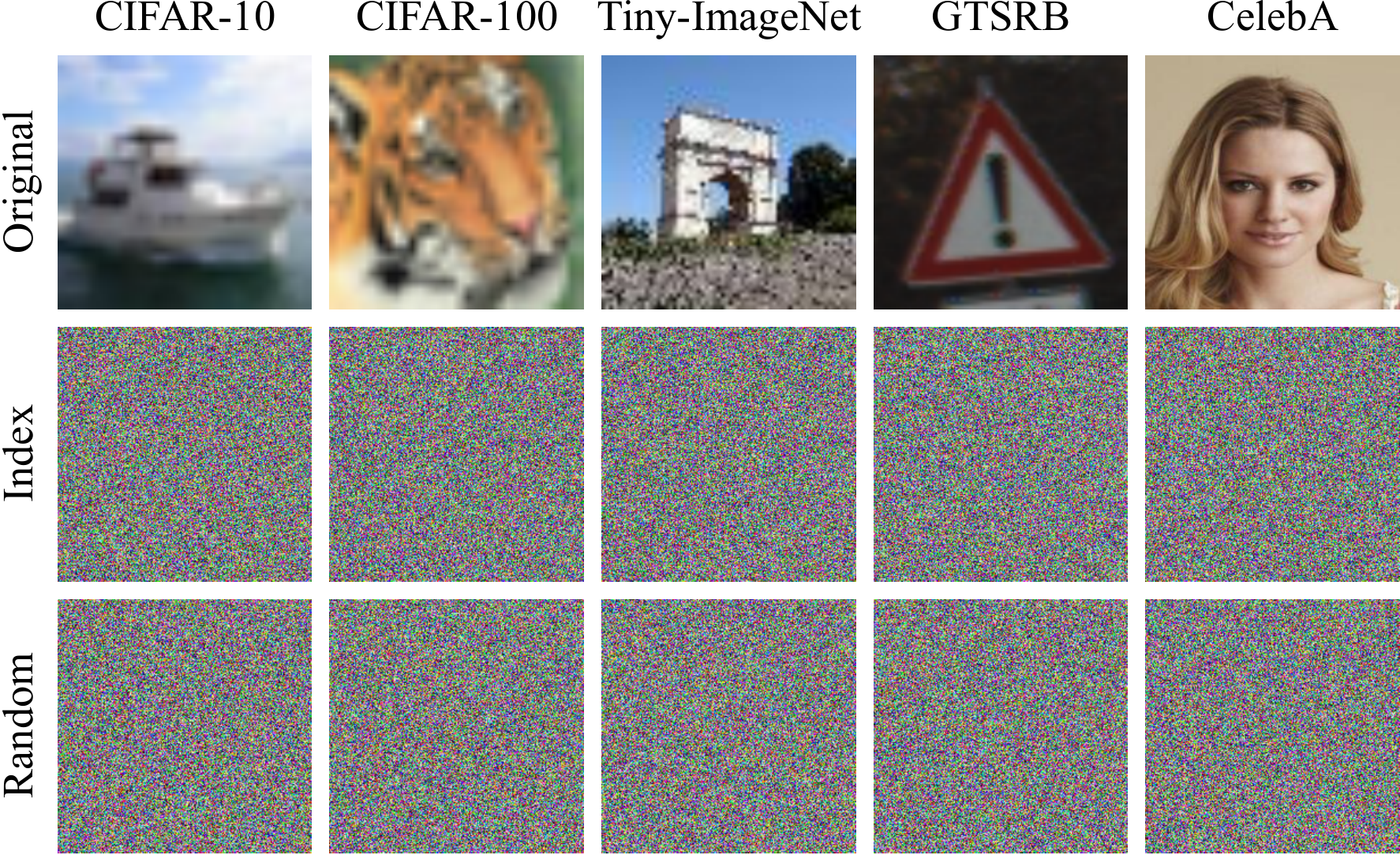}
    \caption{Visual comparison of original images, \net-generated index items, and random images. Each column corresponds to one dataset.}
    \label{fig:viz_cmp}
\end{figure}

\noindent\textbf{Visualization.}
%
%
%
%
Fig. \ref{fig:viz_cmp} compares original images, \net-generated index items, and random images.
For each dataset, the figure shows a randomly selected original image, its corresponding index item, and a random image.
The index items expose no recognizable content from their source images and visually indistinguishable from the random images.


\begin{figure*}
    \centering
    \includegraphics[width=1\linewidth]{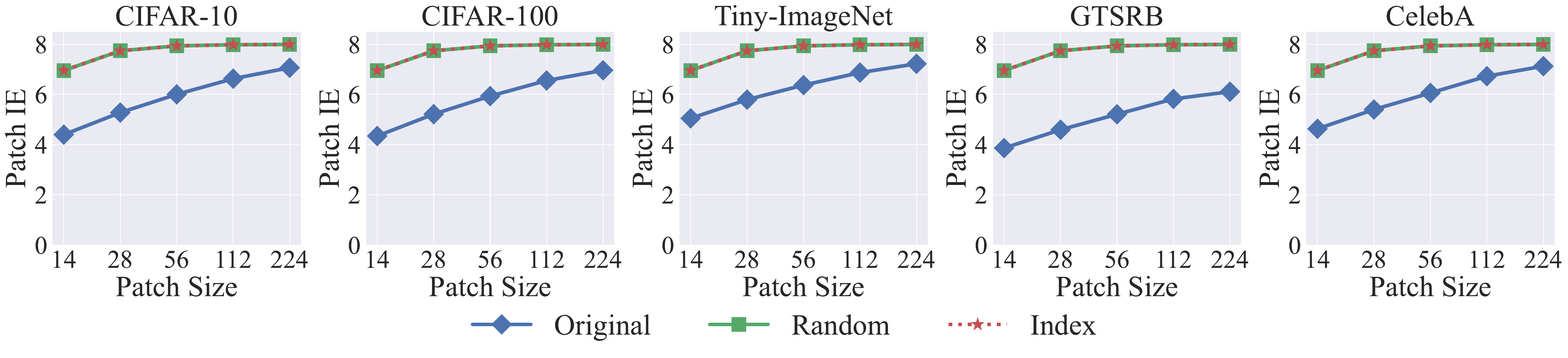}
    \caption{Patch IE of original images, \net-generated index items, and random images across datasets and patch sizes.}
    \label{fig:viz_pie}
\end{figure*}

\noindent\textbf{Patch IE.}
%
%
%
%
Fig. \ref{fig:viz_pie} reports Patch IE for original images, \net-generated index items, and random images over multiple patch sizes. 
Across all datasets and patch sizes, the Patch IE of the index items closely tracks that of random images and is consistently higher than that of original images. 
%

\noindent\textbf{Authorized Accuracy.}
Tables \ref{tab:aa:r18} and \ref{tab:aa:swin2} report AA for ResNet-18
and SwinV2-T, respectively. We compute the accuracy drop as
$\mathrm{AA}_{\mathrm{Original}}-\mathrm{AA}_{\mathrm{Index}}$.
For ResNet-18, the drop ranges from 0.07\% to 3.05\%,
with an average of 1.13\%. For SwinV2-T, it ranges from 0.21\% to 6.13\%, with an average of 2.86\%. 
The largest drop occurs on CIFAR-100 with SwinV2-T, whereas the drop on GTSRB is below 0.25\% for both models. 
Overall, the index items preserve most of the accuracy obtained from the original images, although the utility loss varies by dataset and model.

\begin{table}[H]
\centering
\caption{Authorized accuracy (AA) of ResNet-18. Drop is the absolute decrease from Original to Index.}
{
\begin{tabular}{lrrr}
\toprule
Dataset       & Original & Index $\uparrow$ & Drop $\downarrow$\\
\midrule
CIFAR-10      & 95.75\% & 94.45\% & 1.30\% \\
CIFAR-100     & 79.35\% & 76.30\% & 3.05\% \\
Tiny-ImageNet & 63.39\% & 62.91\% & 0.48\% \\
GTSRB         & 99.42\% & 99.35\% & 0.07\% \\
CelebA        & 80.67\% & 79.94\% & 0.73\% \\
\bottomrule
\end{tabular}}
\label{tab:aa:r18}
\end{table}

\begin{table}[H]
\centering
\caption{Authorized accuracy (AA) of SwinV2-T. Drop is the absolute decrease from Original to Index.}
{
\begin{tabular}{lrrr}
\toprule
Dataset       & Original & Index $\uparrow$ & Drop $\downarrow$\\
\midrule
CIFAR-10      & 95.11\% & 92.23\% & 2.88\% \\
CIFAR-100     & 75.07\% & 68.94\% & 6.13\% \\
Tiny-ImageNet & 63.78\% & 60.03\% & 3.75\% \\
GTSRB         & 99.44\% & 99.23\% & 0.21\% \\
CelebA        & 80.89\% & 79.57\% & 1.32\% \\
\bottomrule
\end{tabular}}
\label{tab:aa:swin2}
\end{table}


\noindent\textbf{Unauthorized accuracy.}
Tables \ref{tab:ua:r18} and \ref{tab:ua:swin2} report UA for ResNet-18 and SwinV2-T, respectively. 
RG denotes the accuracy of uniform random guessing. 
On CIFAR-10, CIFAR-100, and Tiny-ImageNet, UA is close to RG for both architectures, indicating that a model trained on index items is nearly random guessing when applied directly to original images. 
The comparatively higher UA observed on GTSRB and CelebA can be attributed to their imbalanced class distributions, which lead to biased model behaviors.

\begin{table}[H]
\centering
\caption{Unauthorized accuracy (UA) of ResNet-18 across different datasets. RG denotes the accuracy of uniform random guessing.}
{
\begin{tabular}{lrrr}
\toprule
Dataset       & Original & Index $\downarrow$ & RG \\
\midrule
CIFAR-10      & 95.75\% & 10.18\% & 10.00\% \\
CIFAR-100     & 79.35\% &  1.66\% &  1.00\% \\
Tiny-ImageNet & 63.39\% &  0.54\% &  0.50\% \\
GTSRB         & 99.42\% &  8.23\% &  2.33\% \\
CelebA        & 80.67\% & 23.32\% & 12.50\% \\
\bottomrule
\end{tabular}}
\label{tab:ua:r18}
\end{table}

\begin{table}[H]
\centering
\caption{Unauthorized accuracy (UA) of SwinV2-T across different datasets. RG denotes the accuracy of uniform random guessing.}
{
\begin{tabular}{lrrr}
\toprule
Dataset       & Original & Index $\downarrow$ & RG \\
\midrule
CIFAR-10      & 95.11\% & 10.32\% & 10.00\% \\
CIFAR-100     & 75.07\% &  1.55\% &  1.00\% \\
Tiny-ImageNet & 63.78\% &  0.60\% &  0.50\% \\
GTSRB         & 99.44\% &  7.98\% &  2.33\% \\
CelebA        & 80.89\% & 27.16\% & 12.50\% \\
\bottomrule
\end{tabular}}
\label{tab:ua:swin2}
\end{table}


%% file: sections/8-conclusion.tex
\section{Conclusion}\label{sec:conclusion}

In this paper, we have developed a novel framework to enable Secure machine learning queries over encrypted database in cloud Computing. 
\name employs an index-aid approach to achieve security and ML capability simultaneously, while incurring only a slight ML performance degradation.